# Systematic literature review protocol

## Learning-outcomes and teaching-learning process: a Bloom's taxonomy perspective

**Final version: 19/11/2019**

Dr. Samuel Sepúlveda, Dpto. Cs. de la Computación e Informática, Centro de Estudios en Ingeniería de Software, Universidad de La Frontera, Chile.
Dr (c). Mauricio Diéguez, Dpto. Cs. de la Computación e Informática, Centro de Estudios en Ingeniería de Software, Universidad de La Frontera, Chile.
Dr. Gonzalo Farías, Escuela de Ingeniería Eléctrica, Pontificia Universidad Católica de Valparaíso, Chile.
Dra. Cristina Cachero, Dpto. de Lenguajes y Sistemas Informáticos, Universidad de Alicante, España.

## Abstract

***Context***: The importance of defining learning outcomes and the planning stage for a systematic literature review.

***Objective:*** A protocol for carrying out a systematic literature review about the evidence for the tool support for the learning outcomes and the teaching-learning process using Bloom's taxonomy to address it.

***Method***: The definition of a protocol to conduct a systematic literature review according the guidelines of B. Kitchenham.

***Results***: A validated protocol to conduct a systematic literature review.

***Conclusions:*** A proposal for the protocol definition of a systematic literature review about the tool support for the learning outcomes, the teaching-learning process using Bloom's taxonomy was builded it. Initials results show that a more detailed review of the learning outcomes and their alignment with the levels of curricular progress, Training Cycles, and Bloom's Taxonomy should be carried out.

## 1. Introduction

The university teaching-learning process has undergone major changes. Teaching focused on the transmission of content to one focused on the achievement of significant learning by the student.

Nowadays is possible to show a historical context for the improvement of teaching capacities and services declared in the development plan of the Depto. de Cs. de la Computación e Informática in Universidad de La Frontera.

Besides, we believe that there is an edge that has been tackled tangentially and that is related to the importance of defining learning outcomes. The coherence of theses learning outcomes with each other, with the subject, their level in the curriculum, and the profile of the graduate.

To define a subject program, the first element to consider is the set of learning outcomes that the student must show that he can achieve. Based on these, the evaluations that account for their compliance are defined. Then, the methodologies that guarantee compliance are established, and finally, the contents are established in detail. A research opportunity is visualized. It is possible to support the analysis of the curriculum design. An "intelligent" software tool that manages the amount of data, generates statistics and reports to help experts in the process of analysis and decision making.

On the other way, we believe to be in a precious institutional moment. The university is updating its professional training policy (Política de Formación, 2007), the Engineering Faculty considers these elements in its development plan (Plan Estratégico de DEsarrollo, FICA 2017). Finally, all the engineering careers are under a curriculum redesign process.

All these considerations are in the foundations of the research project IF19-0010 and considering the main elements of the active teaching-learning process focused on the student (Manual de Orientaciones, 2018). The goal of this project considers to carry out an analysis and design/development of a support platform to review the learning outcomes present in the subject programs.

The rest of the report is structured as follows. The next section describes the research method to follow, showing the details about the protocol definition for carrying out the systematic literature review. Finally, Section 3 present the conclusions and future work.

## 2. Protocol definition

This section details the protocol definition of a systematic literature review. This protocol considers the three main phases according to the protocol defined by (Kitchenham 2004). These phases are planning, conducting, and reporting.

### 2.1 Planning

The planning considers declare the aim and need for the SLR, and define the research questions that he SLR must answer. Also, this considers the construction of the search string that will allow gathering the relevant papers for the study. Finally, the main considerations about the validation of this protocol are also considered.

#### 2.1.1 Aim and need

This SLR aims to summarize and synthesize the evidence about the learning-outcomes (LOs) and the teaching learning process using Bloom's taxonomy (TxB) to address it. Besides, we are interested in finding tool support for the gathered proposals.

The need is associated on the one hand, with the lack of work that collects evidence on the relationship between learning outcomes and TxB. On the other hand, there is an interest in part of the educational community for the need for tool support that facilitates the development of tasks such as curriculum review and lecture programs.

#### 2.1.2 Research questions (RQs)

***2.1.2.1 Context:*** We present a conceptual background and organizational background.

***i. Conceptual background***

Engineering education has been forced to migrate its traditional approaches (focused on teaching), towards a model that focuses on learning (Capote, 2016), this focus is the student, in such a way to generate new learning opportunities, interactive and collaborative, that provide continuous learning for the student. This change implies that both the curricula of engineering careers and the teachers of the subjects have had to adapt to this trend. Then, the teacher must structure a subject plan that accounts for active and permanent learning for the development of generic and disciplinary skills that an engineer requires.

This plan requires that the LOs that a student must achieve be explained, which, in turn, must derive from the competencies that they wish to develop. According to Jenkins and Unwin (Jenkins and Unwin, 2001), a learning outcome is defined as *"statements that explain what a student will know or be able to do, once the learning process is over*." According to Kennedy (Kennedy, 2006), to write a LO, TxB is frequently used (Bloom, 1975), since it provides a structure and a list of verbs categorized by levels of thinking. The proposal is that each learning result begins with a verb that describes the scope of what is expected of the student regarding the levels proposed by Bloom.

Then, to examine whether or not the student achieved committed learning, all learning outcomes must be evaluated. This requires that there be evaluation instruments and methodological strategies that favor the achievement of the result. Besides, there must be an articulated chain between the competence, the result of learning, its evaluation and the methodological strategy, in such a way to give coherence to the process and facilitate student learning.

***ii. Organizational background***

Teaching in the university field has undergone in recent times by major changes, which consider the transition from a teaching focused on the transmission of content to one focused on the achievement of significant learning by the student.

The University of La Frontera has not been oblivious to these changes, and that is why, from what is defined in its professional training policy, it is considered as a central aspect to design educational material incorporating the concepts of active methodologies focused on student, but also considering the influence of new infrastructure in rooms and laboratories, as well as the availability of new technological elements.

This SLR is framed in the research project IF19-0010[1] that aims the development of EM-RA2[2], a methodological approach for the process of reviewing subject programs and in particular for the analysis of LOs, incorporating a software component that provides semi-automatic support to that analysis. From this, the proposal may be of interest to career directors, managers of the curricular processes and for academics in general, when reviewing and designing the compliance and coherence of the LOs concerning the plan of the subject, degree profiles and competencies of the careers. Likewise, it may be of interest for the teaching analysis units when it comes to having evidence of the work done in updating and redesigning when facing future accreditation processes both institutional and career.

***2.1.2.2 Research questions definition***

According with Kitchenham and Charters, we define a context for the RQs guiding this study (Kitchenham and Charters 2007). The context for the RQs is identifying the impact of tools on ease of use, reliability, and performance on tasks associated to evaluate or generate learning outcomes based on Bloom's taxonomy.

---

[1] Research project supported by Vicerrectoría de Investigación y Postgrado, Universidad de La Frontera.

[2] **E**nfoque **M**etodológico para **R**evisión y **A**nálisis de **R**esultados de **A**prendizaje.

The RQs that drive this SLR and their contribution to the general aim are:

| RQ# | Research Question | Aim |
|---|---|---|
| RQ1 | What is the degree of empirical validation of tools to create or review LOs using TxB? | To examine how each proposal was validated: proofs of concept, case studies, experiments, etc. |
| RQ2 | What is the evidence regarding the level of adoption of tools to create or review LOs using TxB in "real settings"? | To determine how many experiences of their use at universities/institutes/schools/departments/ careers have been reported by the different papers. |
| RQ3 | Who was affected by the use of tools to create or review LO using TxB?<br>RQ3.1 How did the use of tools to create or review LOs using TxB affect them? | To recognize the different actors that were affected by the use of the tools described in the proposals. E.g.: student, teacher, career, faculty, curricular management units, etc.<br>How it affected has to do with the "output" of that affection. E.g. subject programs, curricula, curricular meshes, correspondence matrix, etc. |

***i. Publication questions:*** additionally, a set of publication questions (PQs) has been included to complement the gathered information and characterize the bibliographic and demographic space. This includes the type of venue where the papers were published, amount of papers per year, and where the subject has been more developed.

| PQ# | Publication Question | Aim |
|---|---|---|
| PQ1 | Where the papers had been published?<br>PQ1.1 Who published them? | To determine the distribution of papers by type of venue, i.e., journal, conference or workshop.<br>Besides, to determine what publishers are the most relevant for the studied subject. |
| PQ2 | How the quantity of papers has evolved across the time? | To determine the number of publications per year, in the period 2009-2019. |
| PQ3 | Which are the most active countries? | To determine the most active countries in the subject under study. This considers |
| | PQ3.1 What are the author's affiliations? | the affiliations of all authors as same (don't matter if this is first, second, etc or correspondence author).<br>Besides, to classify the author's affiliations in one of the two categories: academy or industry. This considers the affiliations of all authors as same (don't matter if this is first, second, etc. or correspondence author). |

#### 2.1.2.3 Search string definition

According the steps defined in (Kitchenham and Charters 2007):

- from the RQs ---> keywords
  tools; learning outcomes; Bloom's taxonomy
- keywords ---> synonyms
  tools, tool support, learning outcome, teaching learning, Bloom's taxonomy
- build search string using PICOC (Petticrew and Roberts 2008)
  *population*: an application area, educational systems.
  *intervention*: software tool supporting creation or revision of learning outcomes using TxB.
  *comparison*: N/A.
  *outcomes*: the main outcomes of our RQs are the level of validation and adoption in "real settings"
  *context*: the comparison takes place at academy, the participants are academic and students.

#### i. Search string

("learning outcome" OR "teaching learning") AND "Bloom's taxonomy" AND ("tool" OR "tool support")

#### 2.1.2.4 Protocol validation

Initially, one researcher build the protocol, then, the rest of researchers will must evaluate and discuss about the correctness and completeness for the protocol.

Next, the changes and corrections will be agreed among the researchers and a validated version of the protocol will be compiled.

According to (Kitchenham and Charters 2007), we pretend to evaluate the consistency of the protocol. To do this, we have to answer these questions:

- Are the search string appropriately derived from the RQs?

- Will be the data to be extracted properly address the RQs?
- Is the data analysis procedure appropriate to answer the RQs?

Ideally, the agreed compiled version of the protocol will be sended to an external expert to review it.

## 2.2. Conducting

The conducting phase considers to define the search strategy, the inclusion/exclusion criteria, and the data extraction process. Also, it considers a quality assessment about the gathered evidence and, an initial analysis of the threats to validity of the results.

### 2.2.1 Search strategy

Since the defined data sources include search engines, the strings will be entered sequentially with the combinations of these and adapting it to each search engine as appropriate.

Data sources: according to (Brereton and Kitchenham 2007, Kitchenham and Charters 2007) we consider these sources, that are recognized among the most relevant in SE community.

| **Fuente** | **Link** |
|---|---|
| ACM Digital Library | https://dl.acm.org |
| IEEE Xplore | https://ieeexplore.ieee.org/Xplore/home.jsp |
| IET Digital Library | https://digital-library.theiet.org/content/journals |
| Science Direct | https://www.sciencedirect.com |
| Springer Link | https://www.springer.com/ |
| Wiley Inter Science | https://www.onlinelibrary.wiley.com/search/advanced |

***Time period:*** the search considers the period 2009-2019. This is because in preliminary searches using Scholar Google, we can see that approx. 92% of results concentrate in this period.

***Language***: the selected language will be English. It is considered that this decision, in general, would not become a bias for the SLR, because even though there may be references in other languages that could be discarded, many of these works, given their relevance, will also be published in English, in high impact magazines or conferences.

***Search process:*** this process considers two steps.

- Step 1: automatic search on selected electronic databases.
- Step 2: snowballing process (backward & forward), starting from the final list of selected papers to complement the automatic search (Wohlin 2014).

From the collected papers in step 1, we must consider:

- different versions of the same proposal (expanded works or versions), the criterion will be: keep the last one.
- search and eliminate for duplicated works.

For the remain papers, the researchers will independently review the assigned papers and will decide if these are relevant or not for the SLR only reviewing title, abstract and keywords for the papers.
The set of remain papers will be filtered applying the exclusion criteria (EC).

| EC# | Description |
|---|---|
| EC1 | The paper is not written in English. |
| EC2 | The paper venue is not journal, conference or workshop (it means the paper is not in grey literature). |
| EC3 | The paper was not peer-reviewed. |
| EC4 | The paper is a short paper (less four pages). |
| EC5 | The focus of the paper is not on proposals of tool support for learning out-comes based on Bloom's taxonomy. |
| EC6 | The paper do not show empirical results, at least at the toy example level. |

### 2.2.2 Resolving differences and avoiding bias

- to avoid any potential bias due to a particular researcher examining each paper, we verified that the manner of applying and understanding the exclusion criteria was similar for the researchers involved in the SLR (inter-rater agreement).
- the researchers individually deciding on the inclusion/exclusion of a set of xx papers randomly chosen from those retrieved by a pilot selection.

- a test of concordance based on the *Fleiss' Kappa statistic* will be performed as a means of validation (Gwet 2002).
- if *Kappa* < 0.75, then the test failed, and the criterion must be reviewed to get accordance in its interpretation and application, else if *Kappa >= 0.75,* it suggests that the criteria is clear enough (Fleiss 1981).

Other way to do this task is following the criteria defined by (Petersen, Vakkalanka et al. 2015).

With the final list of selected papers, the snowballing process (backward & forward) will be carried out using these papers as "seeds".

### 2.2.3 Data extraction

For the selected papers, relevant data will be extracted in order to answer the RQs and PQs.

The meta-data collected for each paper: (i) title, (ii) authors (each of them), (iii) publication year, (iv) type of publication and ranking, (v) tools reported, and (vi) results and future work.

Besides, the detailed data for answering each RQ must be registered. All this data will be consolidated in a spreadsheet.

| **Paper#** | **RQ1** | **Evidence RQ1** | **RQ2** | **Ev RQ2** | **RQ3** | **Ev RQ3** | **RQ4** | **Ev RQ4** |
|---|---|---|---|---|---|---|---|---|
| 01 | | | | | | | | |
| 02 | | | | | | | | |
| ... | ... | ... | ... | | | | | ... |
| nn | | | | | | | | |
| **Values** | type of venue (text) | | type of experience (text) | | type of actor and output (text) | | ???? | |
| **Chart** | Bar | | Bar | | Bar | | ??? | |

| **Paper#** | **PQ1** | **PQ2** | **PQ3** |
|---|---|---|---|
| 01 | | | |
| 02 | | | |
| ... | | ... | .... |
| nn | | | |
| **Values** | type of study and publisher (text) | year of publication (integer) | country names (text) |
| **Chart** | Bar | Line | Bar |

Another resource to consolidate data is using a weighted word cloud from the abstracts of the selected paper to show the most relevant concepts. See next image as an example.

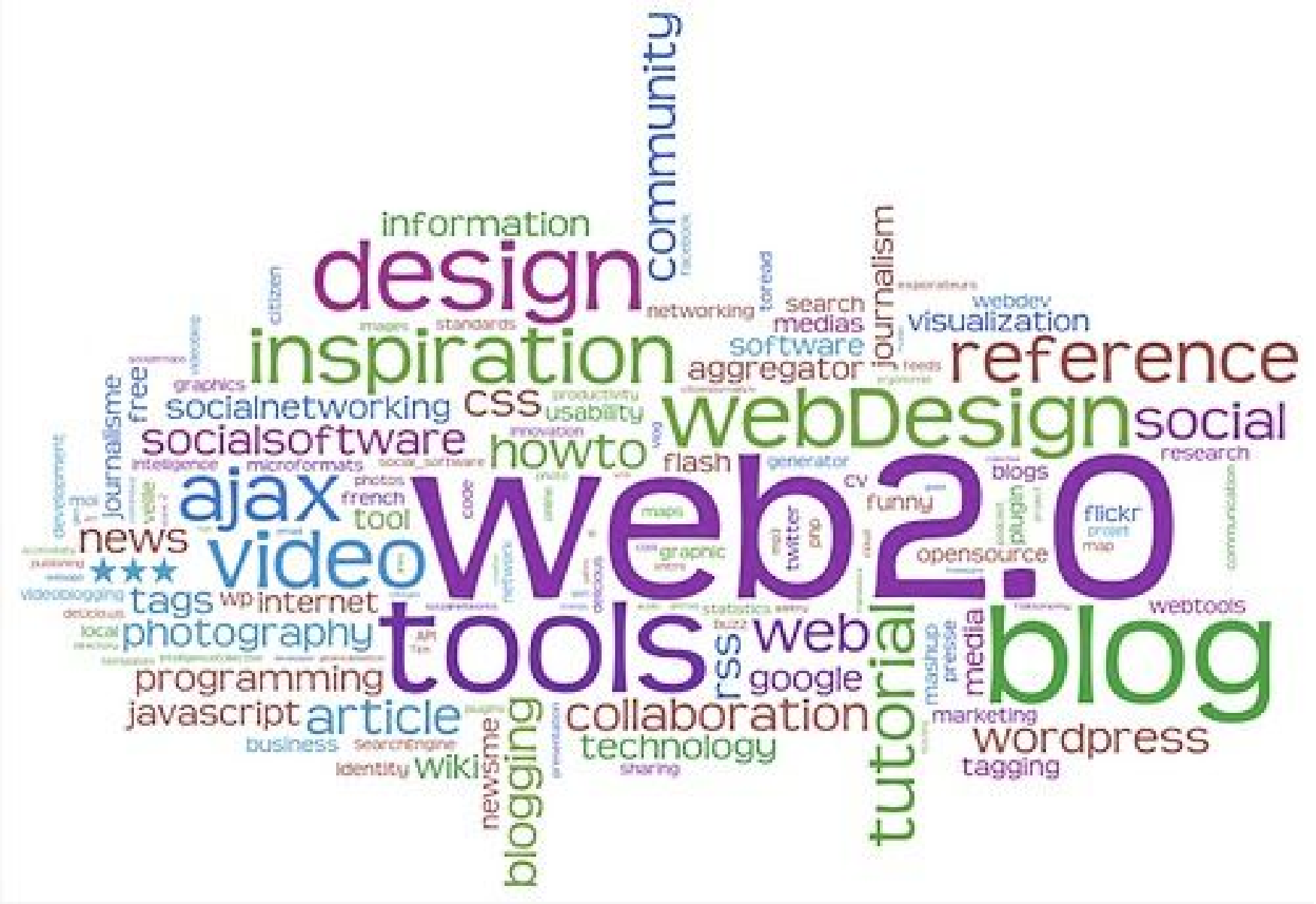

### 2.2.4 Quality assessment (QA)

In order to address bias, external and external validity, each selected paper was subjected to a QA (Kitchenham and Charters 2007). To answer about the QA we will use the criteria proposed by (Dyba and Dingsoyr 2008). These criteria are summarized in five questions that can be answered with *yes/partially/no*. Other papers have already adopted the same criteria, among these are (Chen and Babar 2011, Unterkalmsteiner, Gorschek et al. 2012, Sepúlveda, Cravero et al. 2015). The QA is summarized in the next table.

| **QA#** | **QA question** | **Yes** | **Partially** | **No** |
|---|---|---|---|---|
| QA1 | Is the aim of the research sufficiently explained? | ## (%) | ## (%) | ## (%) |
| QA2 | Is the paper based on research methodology? | ## (%) | ## (%) | ## (%) |
| QA3 | Is there an adequate description of the context in which the research was carried out? | ## (%) | ## (%) | ## (%) |
| QA4 | Are threats to validity taken into consideration? | ## (%) | ## (%) | ## (%) |
| QA5 | Is there a clear statement of findings? | ## (%) | ## (%) | ## (%) |

### 2.2.5 Threats to validity

This sub-phase considers a two-step process to deal with the potential validity threats.

***Step 1:*** Consider the types of validity defined by (Petersen 2015): descriptive validity, theoretical validity, generalizability, interpretive validity.

***Step 2:*** taking care about the common problems about secondary studies reported:
- Possible bias on searching for papers.
- Possible bias in excluding relevant papers.
- Limitations on data extraction from the selected papers
- Limitations on conducting the process and tool support

In a complementary way, as a quality indicator for the report of this SLR, its structure will be validated using the PRISMA Statement[3] (Preferred Reporting Items for Systematic Reviews and Meta-Analyses).

## 2.3. Reporting

To state the relevance of this SLR, the reporting phase considers two steps.
- Step1: publish the reviewed SLR protocol at Arxiv platform[4].
- Step2: publish the SLR results in JCR journal. Initially we consider the following options: IST[5], JSS[6], IEEE Transactions on Education[7].

Adapting the structure recommended by (Kitchenham 2004). The main sections considered are:
1. Introduction: context, motivation, aim, need and structure of the paper.

[3] http://www.prisma- statement.org
[4] https://arxiv.org
[5] https://www.journals.elsevier.com/information-and-software-technology
[6] https://www.journals.elsevier.com/journal-of-systems-and-software
[7] https://ieeexplore.ieee.org/xpl/RecentIssue.jsp?punumber=13

2. Background: main concepts about learning outcomes, TxB and teaching learning process from a technological perspective.
3. Methodology: explain step by step how the protocol will be carried out.
4. Results and discussion: main findings and results to summarize the answers to RQs and PQs, QA analysis and revision of main threats & validity to the paper.
5. Related work: summary of the main related work and RQs answered.
6. Conclusions: conclusions and further lines of research.
7. Acknowledgements: recognize the support for the paper.
8. References: list of bibliographic references, and list of selected papers.
9. Appendix: detailed information (tables and graphs) for complementing sections 3, 4, 5, and 6.

## 3. Conclusions and future work

We presented a proposal for the protocol definition of a systematic literature review to summarize and synthesize the evidence about the tool support for the learning outcomes and the teaching-learning process using Bloom's taxonomy to address it.

The initials results show that a more detailed review of the learning outcomes and their alignment with the levels of curricular progress, Training Cycles, and Bloom's Taxonomy should be carried out.

As future work we pretend to deep in the analysis of the learning outcomes and its relationship with the Bloom's Taxonomy, considering: incorporate the profile and qualification skills, among other elements, and develop a recommender system for the most relevant teaching-learning strategies according to the characteristics of each course.

### *Acknowledgments*

Thanks to Research Project IF19-0010 supported by Vicerrectoría de Investigación y Postgrado, Universidad de La Frontera.